\documentclass[12pt,twoside,titlepage,reqno]{article}
\usepackage{amsmath}
\usepackage{amsthm}
\usepackage{latexsym}
\usepackage{epsf}
\usepackage{graphicx, color}
\begin{document}

\theoremstyle{remark}
\newtheorem{remark}{Remark}[section]
\numberwithin{equation}{section}

\parskip 4pt
\abovedisplayskip 7pt
\belowdisplayskip 7pt

\parindent=24pt

\newcommand{\bn}{{\bf n}}
\newcommand{\bnabla}{{\boldsymbol{\nabla}}}
\newcommand{\bu}{{\bf u}}
\newcommand{\bU}{{\bf U}}
\newcommand{\bv}{{\bf v}}
\newcommand{\bx}{{\bf x}}
\newcommand{\bV}{{\bf V}}
\newcommand{\bz}{{\bf 0}}
\newcommand{\cth}{{\mathcal{T}_h}}
\newcommand{\ct}{{\mathcal{T}}}
\newcommand{\f}{{\bf f}}
\newcommand{\g}{{\bf g}}
\newcommand{\gx}{{\Gamma_{-} (0, \Delta t)}}
\newcommand{\into}{{\displaystyle{\int_{\Omega}}}}
\newcommand{\intG}{{\displaystyle{\int_{\Gamma}}}}
\newcommand{\oo}{{\overline{\Omega}}}
\newcommand{\ox}{{\Omega \times (0, \Delta t)}}
\newcommand{\oxot}{{\Omega \times (0,T)}}
\newcommand{\R}{{I\!\!R}}
\newcommand{\blambda}{{\boldsymbol{\lambda}}}
\newcommand{\bmu}{{\boldsymbol{\mu}}}

\newpage
\thispagestyle{empty} \noindent{\Large\bf Numerical simulation of red blood cell suspensions}
\vskip 1ex \noindent{\Large\bf behind a moving interface in a capillary} 
\vskip 1ex 
\normalsize \noindent{Shihai Zhao and Tsorng-Whay Pan} \vskip 1ex
\noindent{Department of Mathematics, University of Houston, Houston,
Texas  77204-3008, USA}

\vskip 5ex

\noindent{\bf Abstract}: Computational modeling and simulation are presented on the motion of
red blood cells behind a moving interface in a capillary. The methodology is based on an 
immersed boundary method and the skeleton structure of the red blood cell (RBC) membrane
is modeled as a spring network.  The computational domain is moving with either a designated
RBC or an interface in an infinitely long two-dimensional channel with an undisturbed flow field in
front of the domain. The tanking-treading and the inclination angle of a cell in a simple
shear flow are briefly discussed for the validation purpose. We then present the results of the
motion of red blood cells behind a moving interface in a capillary, which show that
the RBCs with higher velocity than the interface speed form a concentrated slug behind
the interface. 


\vskip 2ex \noindent {\bf Key words:}
red blood cells, moving domain, immersed boundary method.

\section{Introduction}

\normalsize

\begin{figure}
\begin{center}
\leavevmode \epsfxsize=2.5 true in \epsffile{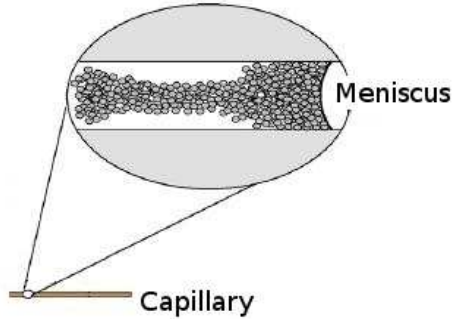}
\end{center}
\vskip -10pt
\caption{Schematics of the BRCs moving behind a meniscus.}\label{fig.1}
\end{figure}

The rheological property of the red blood cells (RBCs) is a key factor of the blood flow characteristics 
at the microchannel level, especially the particulate nature of the blood becomes significant when studying
blood drop through a glass capillary within miniature blood diagnostic kit. The  
penetration of the blood suspension in a perfectly wettable capillary has been analyzed 
in \cite{Chang2005,Chang2006}. The failure of such penetration is attributed 
to three RBCs segregation mechanisms: (i) corner deflection at the entrance, (ii) the intermediate 
deformation-induced radial migration and (iii) shear-induced diffusion within a packed slug at the 
meniscus. The key mechanism responsible for penetration failure is the deformation-induced 
radial migration, which  endows the blood cells with a higher velocity than 
the meniscus to form the concentrated slug behind the meniscus (see Figure \ref{fig.1}).
The results in \cite{Chang2005,Chang2006} shed light on making the smallest microfluidic kit and loading 
microneedle that require the least amount of blood sample.

Nowadays {\it in silico} mathematical modeling and numerical study of RBC rheology  
have attracted growing interest (see, e.g., \cite{Cristini, Pozrikidis2003}). 
The immersed boundary method developed by Peskin, e.g, \cite{Peskin1977, Peskin1980, Peskin2002},
has been one of the popular methodologies for numerically studying the RBC rheology due its distinguish 
features in dealing with the problem of fluid flow interacting with a flexible fluid/structure interface.
For example, in \cite{Eggleton1998}-\cite{Shi2012c}, 
immersed boundary methods have been combined with different RBC membrane models to simulate 
the motion of RBCs and vesicles in fluid flow.  In \cite{Shi2012a,Shi2012b,Shi2012c},
we have successfully combined an immersed boundary method with a spring model developed in \cite{Tsubota2006} 
to simulate the motion of RBCs in shear flows and Poiseuille flows. In this paper we have generalized
the aforementioned methodology to simulate the RBCs  aggregation behind a moving interface 
considered in \cite{Chang2005,Chang2006} by having the computational 
domain moving with an interface in an infinitely long two-dimensional channel with an undisturbed 
flow field in
front of the domain  since the typical periodic boundary condition in the direction of the channel 
wall is not well suited anymore.  To mimic the motion of the RBCs behind a meniscus in a capillary, 
we have considered a flat interface moving with a given constant speed in this paper. The simulating results 
of the motion of red blood cells behind a moving interface show that
the RBCs with higher velocity than the interface speed form the concentrated slug behind
the interface, which resembles the motion of the RBCs observed in  \cite{Chang2005,Chang2006}.
The structure of this paper is as follows: We discuss the elastic 
spring model and numerical methods in Section 2. In Section 3, the tanking-treading and the inclination 
angle of a cell in a simple shear flow are briefly discussed for the validation purpose.
We then present the results of the motion of red blood cells behind a moving interface in a capillary.
The conclusions are summarized in Section 4.

\section{Models and methods}\label{sec:2}

\begin{figure}
\begin{center}
\leavevmode \epsfxsize=5.5 true in \epsffile{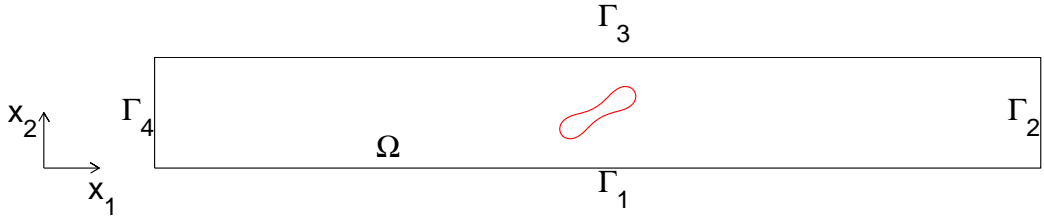}
\end{center}
\caption{An example of the computational domain with a cell.} \label{fig.2}
\end{figure}

Let $\Omega$ be a bounded rectangular domain filled with blood plasma which 
is incompressible, Newtonian, and contains RBCs with the 
viscosity of the cytoplasm  same as that of the blood plasma (see Figure \ref{fig.2}). 
For some $T>0$, the governing equations for the fluid-cell system are 
\begin{equation}
\label{eqn:2.1} \rho \displaystyle \left[ \frac{ \partial {\bf u}}
{\partial t} + {\bf u} \cdot {\boldsymbol \nabla} {\bf u}
\right]=-{\boldsymbol \nabla} p + \mu \Delta {\bf u} +{\bf f} 
\ \text{in}  \enspace \Omega, \ t \in (0,T),
\end{equation}
\begin{equation}
\label{eqn:2.2} {\boldsymbol \nabla} \cdot {\bf u} =0
\enspace  \text{in} \enspace \Omega, \enspace t \in (0,T)
\end{equation}
where ${\bf u}$ and $p$ are the fluid velocity and
pressure, respectively,   $\rho$ is the fluid density, and $\mu$ is
the fluid viscosity, which is assumed to be constant for the entire
computational domain. In (\ref{eqn:2.1}), ${\bf f}$ is a body force  which  
accounts for the force acting on the fluid/cell interface.
Equations (\ref{eqn:2.1}) and (\ref{eqn:2.2}) are completed by the following 
boundary and initial conditions:
\begin{eqnarray}
&& \label{eqn:2.3a} {\bf u} ={\bf g}_0 \enspace \text{on} \enspace \Gamma_d,\\
&& \label{eqn:2.3b} \mu \dfrac{{\partial\bf u}}{\partial\bf n} -{\bf n}p ={\bf 0} \enspace \text{on} \enspace \Gamma_n,\\
&& \label{eqn:2.4} {\bf u}(0) ={\bf u}_{0}
\end{eqnarray}
where the domain $\Omega$ is taken from an infinitely long 
channel  with its boundary denoted by $\displaystyle\Gamma= \cup_{i=1}^4 \Gamma_i$.
In the simulations, we have considered two types of boundary conditions: (i)  $\Gamma_n=\emptyset$ and
$\Gamma_d=\Gamma$,  (ii) $\Gamma_n=\Gamma_4$ and $\Gamma_d=\Gamma_1\cup \Gamma_2 \cup \Gamma_3$
with ${\bf g}_0$ having the profile of either Poiseuille flow or simple shear flow on $\Gamma_d$

\subsection{Elastic spring model for the RBC membrane}

\begin{figure}
\begin{center}
\leavevmode \epsfxsize=2.0 true in \epsffile{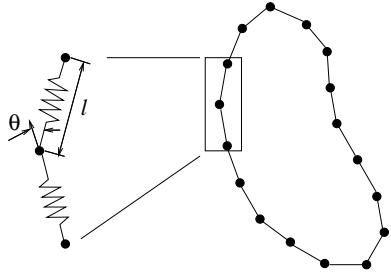}
\end{center}
\caption{The elastic spring model of the RBC membrane} \label{fig.3}
\end{figure}

A two-dimensional elastic
spring model used in \cite{Tsubota2006} is considered in this paper to
describe the deformable behavior of the RBCs. Based on this model,
the RBC membrane can be viewed as membrane particles connecting with
the neighboring membrane particles by springs, as shown in
Figure \ref{fig.3}. Elastic energy stores in the spring due to the
change of the length $l$ of the spring with respected to its
reference length $l_{0}$ and the change in angle $\theta$ between
two neighboring springs. The total elastic energy of the RBC
membrane, $E=E_{l}+E_{b}$, is the sum of the total elastic energy
for stretch/compression and the total energy for bending which, in
particular, are
\begin{equation}
\label{eqn:2.5}
E_{l}=\frac{k_{l}}{2}\sum_{i=1}^{N}(\frac{l_{i}-l_{0}}{l_{0}})^{2}
\end{equation}
and
\begin{equation}
\label{eqn:2.6}
E_{b}=\frac{k_{b}}{2}\sum_{i=1}^{N}tan^{2}(\theta_{i}/2).
\end{equation}
In equations (\ref{eqn:2.5}) and (\ref{eqn:2.6}), $N$ is the total number of
the spring elements, and $k_{l}$ and $k_{b}$ are spring constants
for changes in length and bending angle, respectively.  

\begin{remark} \label{rem:2.1}
In the process of creating the initial shape of RBCs described in \cite{Tsubota2006}, 
the RBC is assumed to be a circle of radius $R_0=2.8 \ \mu m$ initially. 
The circle is discretized into $N=76$ membrane particles  so that $76$ 
springs are formed by connecting the neighboring particles. The shape 
change is stimulated by reducing the total area of the circle
through a penalty function
\begin{equation}
\label{eqn:3.1} \Gamma_{s}=\frac{k_{s}}{2}(\frac{s-s_{e}}{s_{e}})^{2}
\end{equation}
where $s$ and $s_{e}$ are the time dependent area of the RBC and the
equilibrium area of the RBC, respectively, and the total energy is modified as $E+\Gamma_{s}$. Based on the
principle of virtual work the force acting on the
$i$th membrane particle now is
\begin{equation}
\label{eqn:3.2} {\bf F}_{i}=-\frac{\partial(E+\Gamma_{s})}{\partial
{\bf r}_{i}}
\end{equation}
where ${\bf r}_{i}$ is the position of the $i$th membrane
particle.
When the area is reduced, each RBC membrane particle moves on the basis of
the following equation of motion:
\begin{equation}
\label{eqn:3.3}
m\ddot{{\bf r}_{i}}+\gamma\dot{{\bf r}_{i}}={\bf F}_{i}
\end{equation}
Here, $\dot{()}$ denotes the time derivative; $m$ and $\gamma$
represent the membrane particle mass and the membrane viscosity of the RBC. 
The position ${\bf r}_{i}$ of the $i$th membrane particle is solved by discretizing 
(\ref{eqn:3.3}) via a second order finite difference method. The total 
energy stored in the membrane decreases as the time elapses.  The final shape 
of the RBC is obtained as the total elastic energy is minimized (please see \cite{Pan2009b}).
The area of the final shape has less than $0.001\%$ difference from the given equilibrium area $s_e$ 
and the length of the perimeter of the final shape has less than $0.005\%$ difference from the 
circumference of the initial circle. The reduced area of a RBC in this paper is defined 
by $s^{*}=s_e/\pi R_0^2$.
\end{remark}

\begin{remark}
When simulating the case involving a moving interface, we have applied
a repulsive force to prevent the overlapping between cell and wall. The  repulsive force
is obtained from the following Morse potential (e.g., see \cite{Alexeev2006})
\begin{equation*}
\phi(d)=k_r(1-e^{-(d-d_0)})^2
\end{equation*}
where the parameter $d$ is the shortest distance between the membrane particle and the wall 
and $d_0$ is the range of the  repulsive force (when the distance $d$
is greater than $d_0$, there is no repulsive force). The parameter $k_r$ is a constant for the
strength of the potential. 
\end{remark}

\subsection{Immersed boundary method}

The immersed boundary method developed by Peskin, e.g, \cite{Peskin1977, Peskin1980, Peskin2002},
is employed in this study because of its distinguish features in
dealing with the problem of fluid flow interacting with a flexible
fluid/structure interface. Over the years, it has demonstrated its
capability in study of computational fluid dynamics including blood
flow. Based on the method, the boundary of the deformable structure
is discretized spatially into a set of boundary nodes. The force
located at the immersed boundary node ${\bf r}_i=(r_{i,1},r_{i,2})$ affects the
nearby fluid mesh nodes ${\bf x}=(x_1,x_2)$ through a 2D discrete
$\delta$-function $D_{h}({\bf x}-{\bf r}_i)$:
\begin{equation}
\label{eqn:2.8} {\bf f}({\bf x})=\sum {\bf F}_i  D_{h}({\bf x}-{\bf r}_i) \ \ for \
\lvert {\bf x}-{\bf r}_i \rvert \le 2h,
\end{equation}
where $h$ is the uniform finite element mesh size and
\begin{equation}
\label{eqn:2.9}
D_{h}({\bf x}-{\bf r}_i)=\delta_{h}(x_{1}-r_{i,1})
\delta_{h}(x_{2}-r_{i,2})
\end{equation}
with the 1D discrete $\delta$-functions being
\begin{equation}
\label{eqn:2.10} 
\delta_{h}(z)=\begin{cases} 
\frac{1}{8h}\left(3-{2|z|}/{h}+\sqrt{1+{4|z|}/{h}-4({|z|}/{h})^2}\right), &   \ \lvert z \rvert \le h, \\
\frac{1}{8h}\left(5-{2|z|}/{h}-\sqrt{-7+{12|z|}/{h}-4({|z|}/{h})^2}\right), &  \ h \le \ \lvert z \rvert \le 2h, \\
0, & \  otherwise.
\end{cases}
\end{equation}

The velocity of the immersed boundary node ${\bf r}_i$ is also
affected by the surrounding fluid and therefore is enforced by
summing the velocities at the nearby fluid mesh nodes ${\bf x}$
weighted by the same discrete $\delta$-function:
\begin{equation}
\label{eqn:2.11} {\bf U}({\bf r}_i)=\sum h^{2} {\bf u(x)}
D_{h}({\bf x}-{\bf r}_i) \ \ for \ \lvert
{\bf x}-{\bf x}_i \rvert \le 2h.
\end{equation}
After each time step, the position of the immersed boundary node is
updated by
\begin{equation}
\label{eqn:2.12} {\bf r}_i^{n+1}={\bf r}_i^n+\Delta{t}
{\bf U}({\bf r}_{i}^n).
\end{equation}

\subsection{Space approximation and time discretization}

Concerning the finite element based {\it space approximation} of $\{\bu, p\}$ in problem 
(\ref{eqn:2.1})-(\ref{eqn:2.4}), we use the $P_1$-$iso$-$P_2$ and $P_1$
finite element approximation (e.g., see \cite{Glowinskibook} (Chapter 5)).
For a rectangular computational domain $\Omega \subset R^{2}$, let $\ct_{h}$ be a
finite element triangulation of $\overline{\Omega}$ for velocity and
$\ct_{2h}$ a twice coarser triangulation for pressure where 
$h$ is a space discretization step.
We introduce the finite
dimensional spaces:

\begin{eqnarray*}
&&W_h=\{{\bf v}_{h}|{\bf v}_{h}\in C^{0}(\overline{\Omega})^{2}, {\bf v}_{h}|_{T}\in P_{1}\times
P_{1}, \forall T \in \ct_{h}\}\\
&&W_{0h}=\{{\bf v}_{h}|{\bf v}_{h}\in W_h, {\bf v}_{h}={\bf 0} \ \text{on}  \ \Gamma_d \},\\
&&L_{h}^{2}=\{q_{h}|q_{h}\in C^{0}(\overline{\Omega}), q_{h}|_{T}\in
P_{1}, \forall T \in \ct_{2h} \ \},\\
&&L_{h,0}^{2}=\{q_{h}|q_{h}\in L_{h}^{2}, \int_{\Omega} q_h\, d\bx=0 \}
\end{eqnarray*}
where $P_{1}$ is the space of polynomials in two
variables of degree $\leq 1$.  We apply the {\it Lie's scheme} \cite{Chorin, Glowinskibook} with
the above finite elements to equations (\ref{eqn:2.1})-(\ref{eqn:2.4}) with the backward
Euler method in time for some subproblems and obtain the following sequence of 
fractional step subproblems (some of the subscripts $h$ have
been dropped):

${\bf u}^{0}={\bf u}_{0}$ is given; 
for $n\geq 0$, ${\bf u}^{n}$ being known, we compute the approximate solution
via the following fractional steps:

\begin{enumerate}
\item Update the position of the membrane by (\ref{eqn:2.11}) and (\ref{eqn:2.12}) 
and then compute the force  ${\bf f}^n$ based on the fluid/cell interface 
by (\ref{eqn:3.2}) and (\ref{eqn:2.8}). 

\item Solve
\begin{equation}
\left\{
\begin{array}{ll} \displaystyle
\int _{\Omega}\frac{\partial \bu(t)}{\partial t} \cdot \bv
d\bx+\int _{\Omega}(\bu^{n}\cdot \bnabla)\bu(t)\cdot \bv
d\bx=0, \ \enspace \text{on}\, \enspace (t^{n}, t^{n+1}), \enspace \forall \bv
\in W_{0h}^-,\\
\bu(t^{n})=\bu^{n},\\
\bu(t)\in W_h, \ \bu(t)={\bf g}_{0,h} \ \enspace \text{on}\, \enspace \Gamma^- \times (t^{n},t^{n+1}),
\end{array}\right.\label{eqn:2.17}\\
\end{equation}
and set $\bu^{n+2/3}=\bu(t^{n+1})$. 

\item Finally solve
\begin{equation}\
\left\{
\begin{array}{ll}\displaystyle
\rho \int _{\Omega} \frac{\bu^{n+1}-\bu^{n+2/3}}{\triangle t} \cdot
\bv d\bx +\mu \int _{\Omega} \bnabla \bu^{n+1} \boldsymbol{:} 
\bnabla \bv d\bx  \\
\hskip 20pt -\displaystyle\int_{\Omega} p^{n+1} (\bnabla \cdot \bv) d\bx \displaystyle
=\int _{\Omega} {\bf f}^n \cdot \bv d\bx, \enspace \forall \bv \in W_{0h},\\
\displaystyle \int _{\Omega} q\bnabla \cdot \bu^{n+1}d\bx =0,
\enspace \forall q \in L_{h}^{2},\\
\bu^{n+1}\in W_h,\ \bu(t)={\bf g}_{0,h} \ \text{on} \ \Gamma_d; \  p^{n+1}\in L_{h}^{2} \
(p^{n+1}\in L_{h,0}^{2} \ \text{if} \ \Gamma_n=\emptyset).
\end{array}\right.\label{eqn:2.16}
\end{equation}
\end{enumerate}
In eq. (\ref{eqn:2.17}), we have $\Gamma^{-}=\{\bx|\bx \in \Gamma$, ${\bf g}_{0,h}(\bx)\cdot {\bf n}(\bx)<0$ $\}$ 
and $W_{0h}^-=\{{\bf v}_{h}|{\bf v}_{h}\in W_h, {\bf v}_{h}={\bf 0} \ \text{on}  \ \Gamma^- \}$.
The quasi-Stokes problem (\ref{eqn:2.16}) is solved by a preconditioned
conjugate gradient method (see, e.g., \cite{Glowinskibook}).
The subproblem (\ref{eqn:2.17}) is an advection type subproblem. It is
solved by a wave-like equation method, which is described in detail
in \cite{Glowinskiwave} and \cite{Dean1998}.

\begin{remark} 
In simulations, the computational domain $\Omega$ moves to the right with either 
the mass center of a RBC or the interface (see, e.g., \cite{Hu1992, Pan2002c} and references 
therein for adjusting the computational domain according to the position of the particle). 
Due to the use of structured and uniform mesh in our simulations, it is relatively easy
to have the computational domain moving with a designated cell.
Generally when the mass center of a RBC moves to the right in an infinitely long channel,
we add one vertical grid line to the right end of the computational domain if the cell mass 
center crosses one vertical grid line after we predict its new position
and at the same time we drop one vertical grid line at the left end of the 
computational domain. In the mean time at these new grid points added at the right end, we assign the
values of velocity field according to either Poiseuille flow or simple shear flow
depending on the test cases. When following an interface moving to the right with a constant
speed, we have applied the same strategy.
\end{remark}

\section{Numerical results and discussion}

\subsection{Tank-treading of a single cell in shear flow}\label{sec.3.1}

\begin{figure}[ht]
\begin{center}
\leavevmode
\epsfxsize=3.0in \epsffile{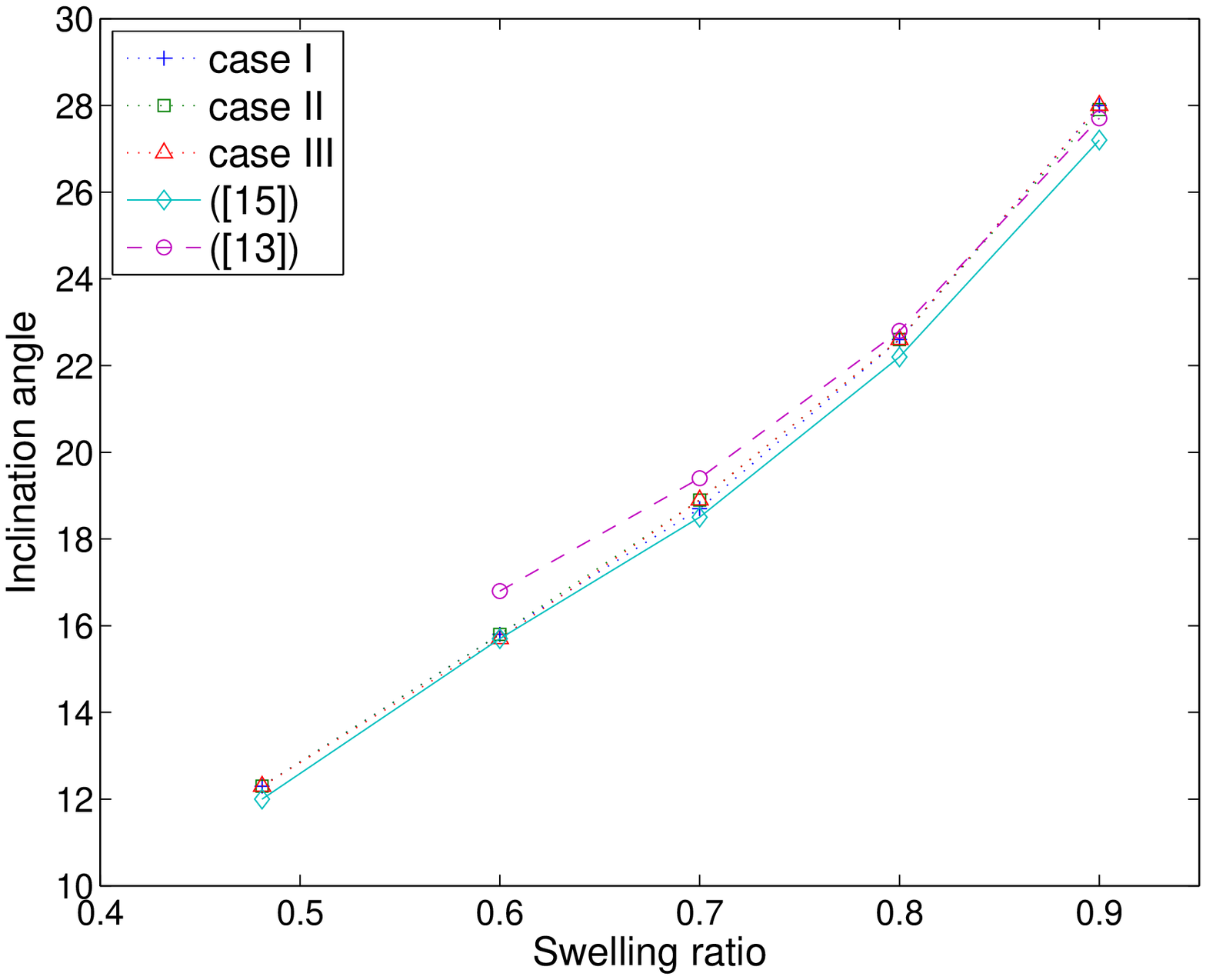}
\hskip 5pt
\epsfxsize=3.0in \epsffile{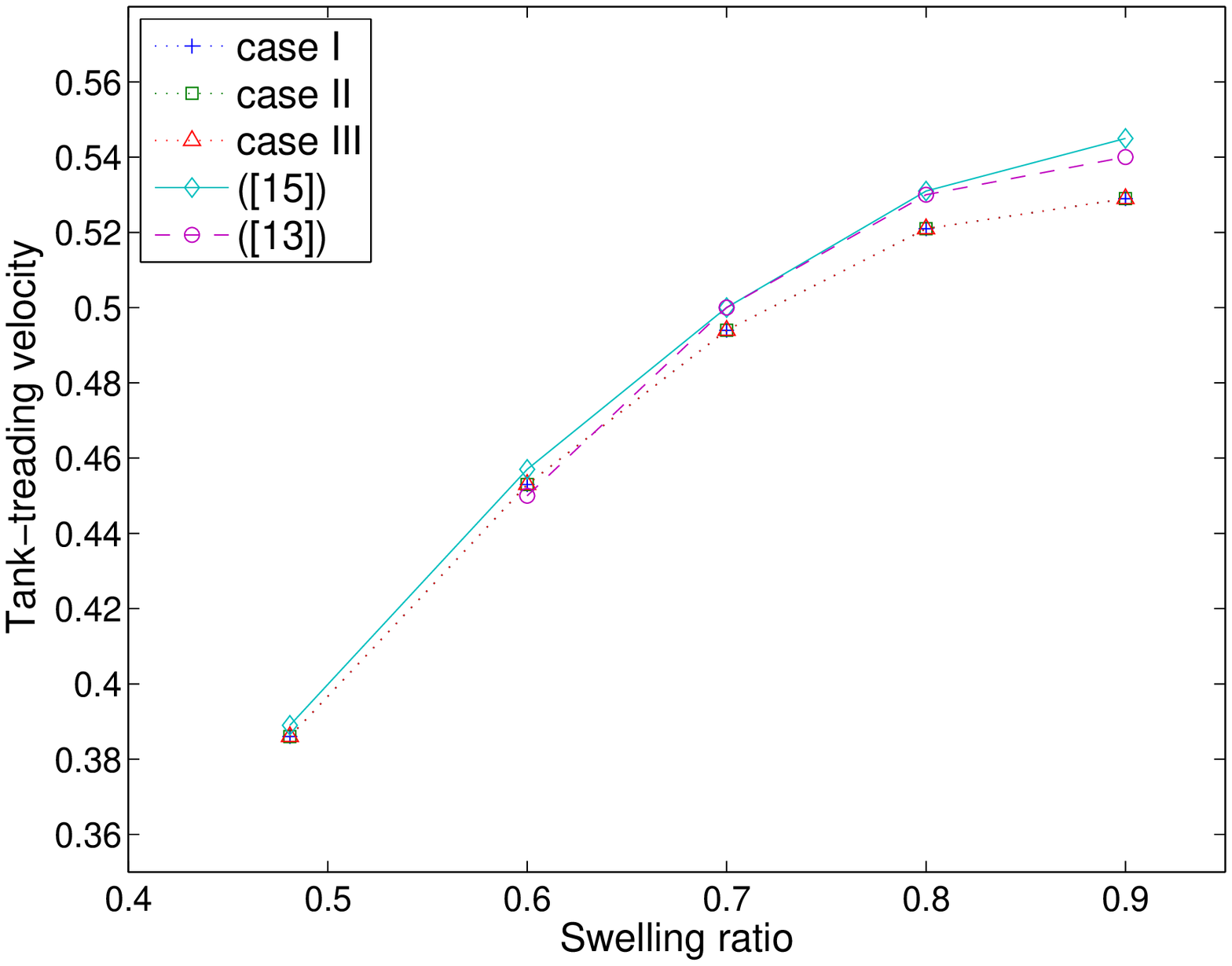}
\end{center}
\caption{(Color online).  Steady inclination angle versus the cell swelling ratio (left) and membrane tank-treading velocity (scaled by $\gamma R_0/2$) versus the cell swelling ratio (right) in comparison to Shi. et al.\cite{Shi2012a} and Kaoui. et al.\cite{Kaoui2011} in different cases. Case I: $112 \rm{\mu m }\times 7\rm{\mu m}$ domain with Dirichlet boundary conditions, Case II: $80\rm{\mu m} \times 7\rm{\mu m}$ domain with Dirichlet boundary conditions, Case III: $80\rm{\mu m} \times 7\rm{\mu m}$ domain with Neumann inflow condition and Dirichlet outflow condition.} \label{fig.4}
\end{figure}

\begin{figure}
\begin{center}
\leavevmode
\epsfxsize=5.5in \epsffile{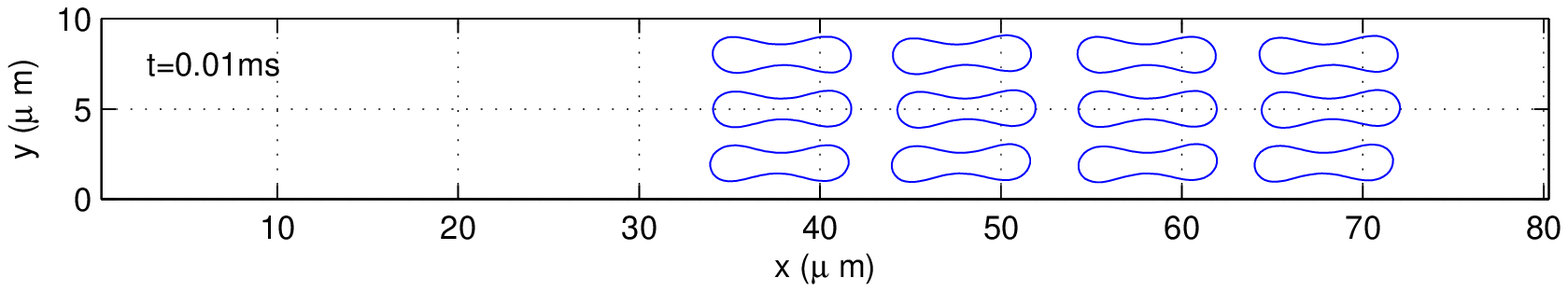}\\
\epsfxsize=5.5in \epsffile{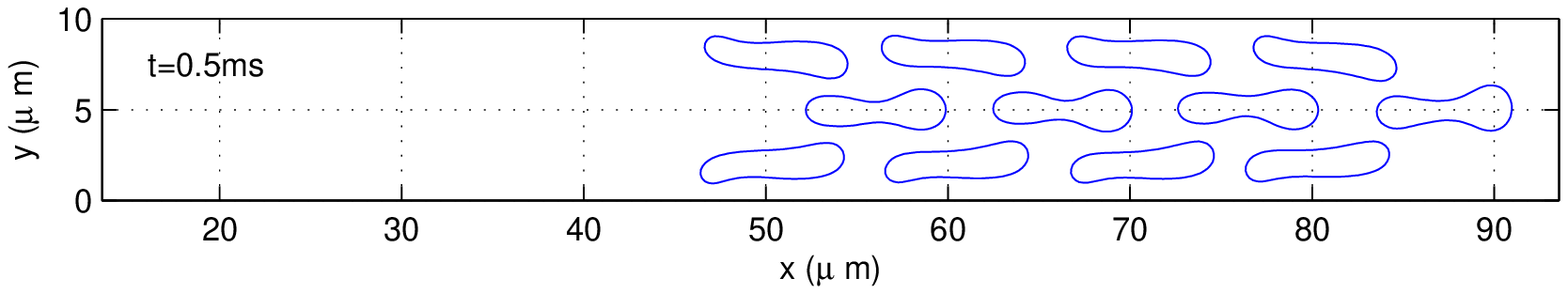}\\
\epsfxsize=5.5in \epsffile{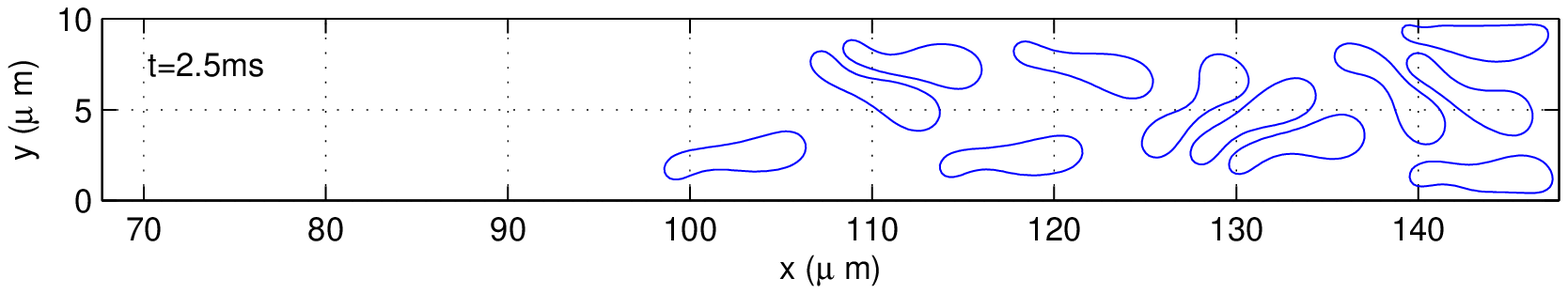}\\
\epsfxsize=5.5in \epsffile{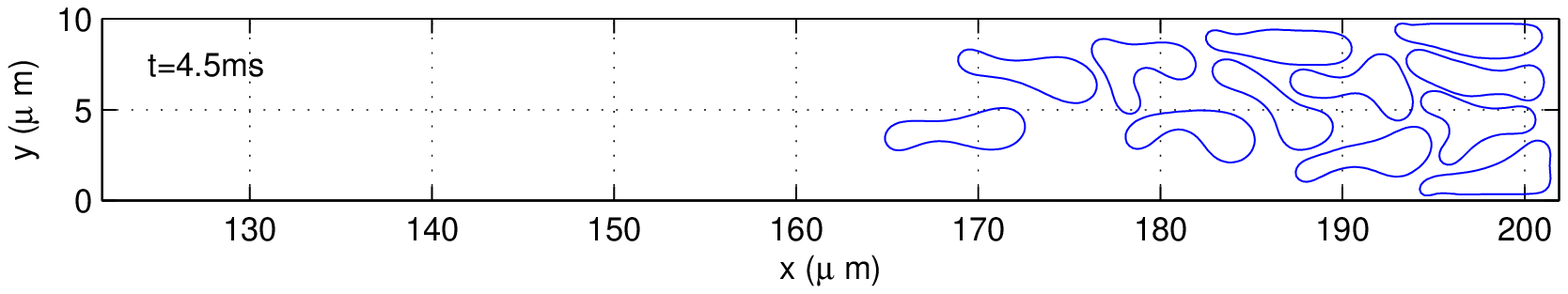}\\
\epsfxsize=5.5in \epsffile{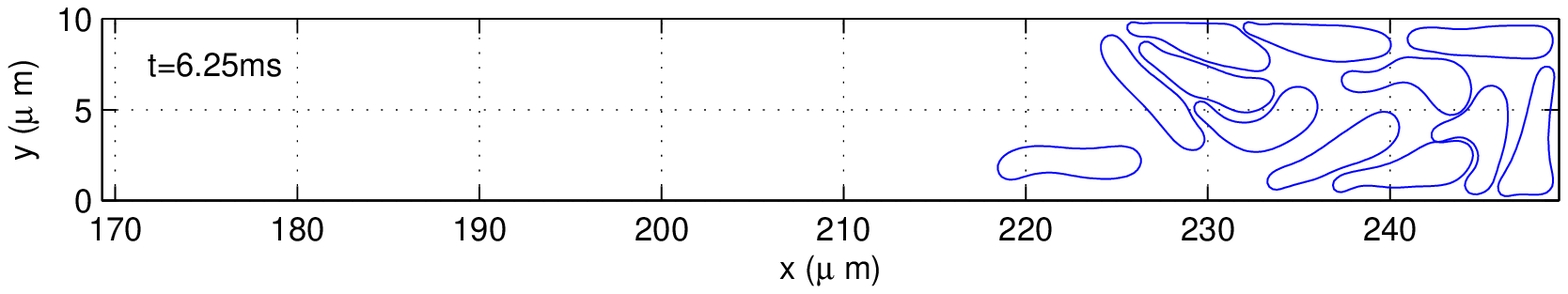}\\
\vskip 5pt
\hskip 0.44in\epsfxsize=5.25in \epsffile{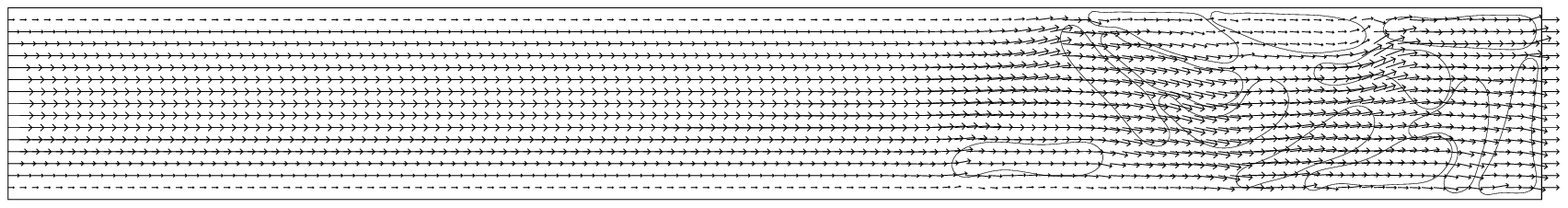}\\
\end{center}
\caption{(Color online). The positions of 12 cells in a capillary behind a moving interface at $t=$
0.01, 0.5, 2.5, 4.5 and 6.25 ms and the velocity field with 12 cells at $t=$6.25 ms (from top to bottom).} \label{fig.5}
\end{figure}

\begin{figure}
\begin{center}
\leavevmode
\epsfxsize=6.in \epsffile{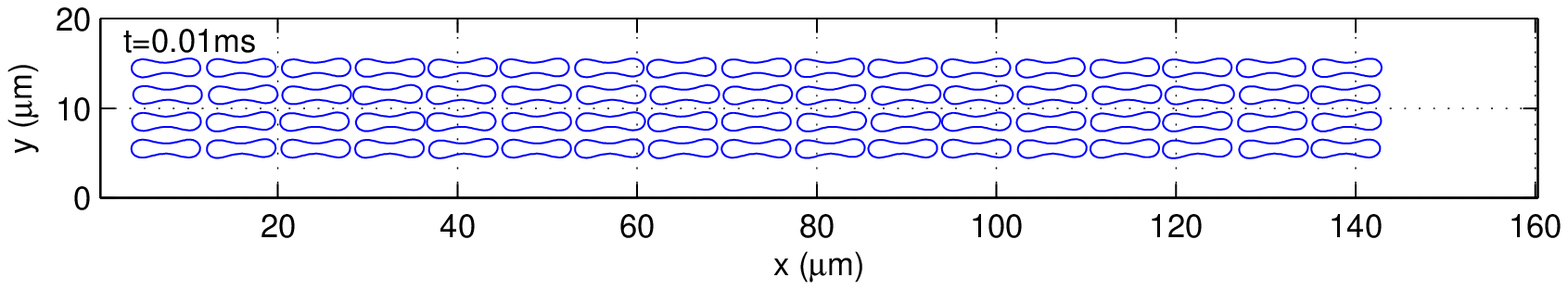}\\
\epsfxsize=6.in \epsffile{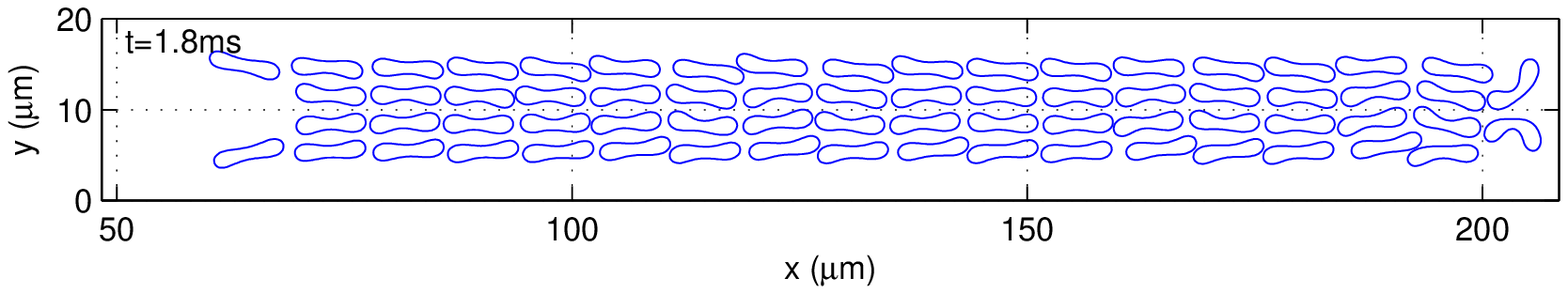}\\
\epsfxsize=6.in \epsffile{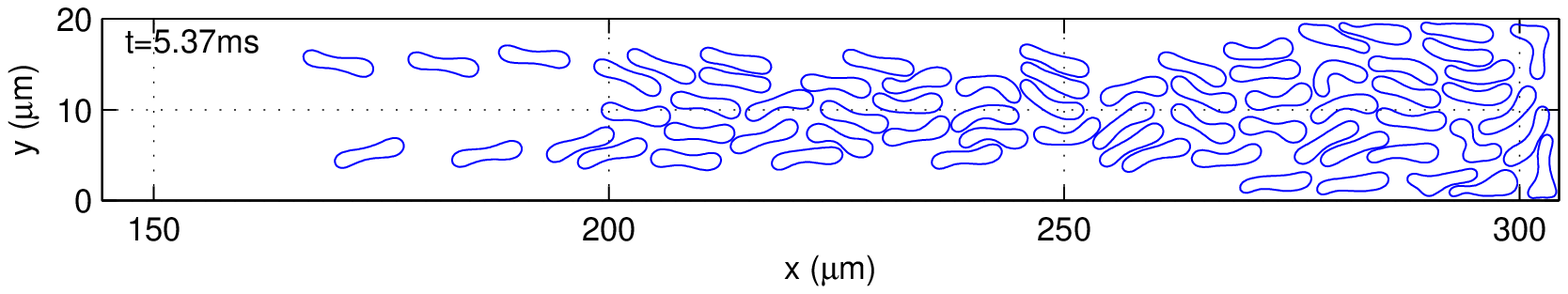}\\
\hskip 0.39in\epsfxsize=5.57in \epsffile{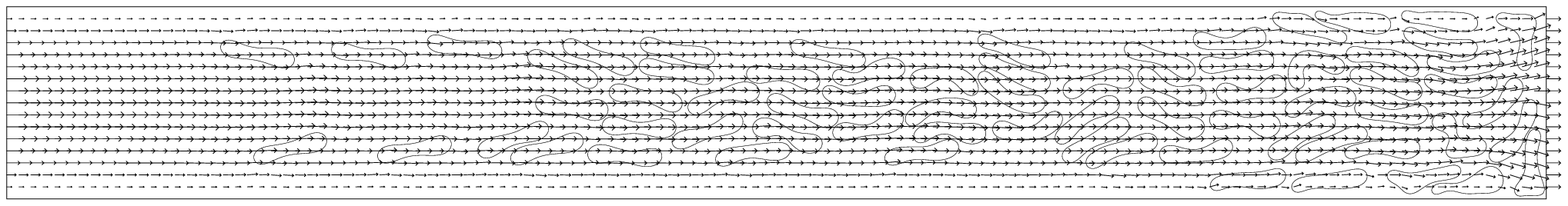}\\
\vskip 5pt
\epsfxsize=6.in \epsffile{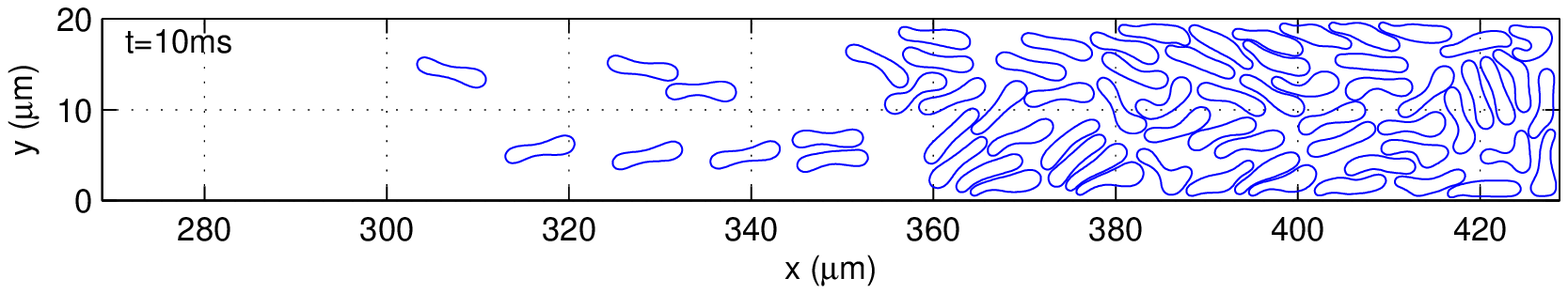}\\
\hskip 0.39in\epsfxsize=5.57in \epsffile{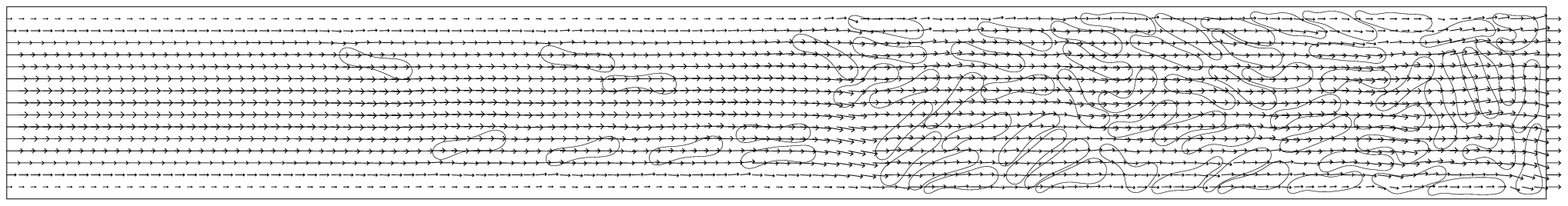}\\
\end{center}
\caption{(Color online). The position of 68 cells in a capillary behind a moving interface at $t=$
0.01, 1.8, 5.37 and 10 ms and the velocity field with 68 cells at $t=$ 5.37 and 10 ms (from top to bottom).} \label{fig.6}
\end{figure}

We have first validated the computational methodology with two types of boundary conditions
discussed in Section \ref{sec:2} by comparing the inclination angle and the tank-treading
frequency of a single RBC in shear flow. Here are the parameters used in the simulations:
The values of parameters for modeling cells are same with \cite{Shi2012a,Shi2012b,Shi2012c} as follows:
The bending constant is $k_{b}=5\times 10^{-10}\rm{N \cdot m}$, the spring constant is
$k_{l}=5\times10^{-8}\rm{N \cdot m}$, the penalty coefficient is $k_s= 10^{-5}\rm{N \cdot m}$, 
the repulsive force coefficient is $k_r= 10^{-9}\rm{N \cdot m}$, and the range of the repulsive force
is $d_0=2h$ where $h$ is the mesh size for the flow velocity field.
The cells are suspended in blood plasma which has a density $\rho=1.00\rm{g/cm^{3}}$ and a dynamical
viscosity  $\mu=0.012 \rm{g/(cm\cdot s)}$. The viscosity ratio which describes the viscosity contrast
of the inner and outer fluid of the RBC membrane is fixed at 1.0. The dimensions of the computational 
domain are $112 \rm{\mu m }\times 7\rm{\mu m}$ and $80\rm{\mu m} \times 7\rm{\mu m}$. 
Then the degree of confinement ($2R_0/H$) is 0.8 where $H$ is the height of the channel.
The grid resolution for the computational domain is 80 grid points per 10$\rm{\mu m}$.
The time step $\Delta t$ is $1 \times 10^{-5} \rm{ms}$.  The initial position of the mass center of the cells
are (56, 3.5) and (40, 3.5) for the longer domain and the shorter domain, respectively.  
To have a shear flow, a Couette flow driven by two walls at the top and bottom which have the same speed  $U/2$
but move in directions opposite to each other is applied to the suspension, where the speed $U$ is given by $U=\gamma*H$ 
with a given shear rate $\gamma$. The shear rate used in the simulation is $\gamma=275$/s.
The steady inclination angles of the tank-treading for
four values of $s^*$=0.6, 0.7, 0.8 and 0.9 are presented in Figure \ref{fig.4},
which show the very good agreement with the lattice-Boltzmann simulation results in \cite{Kaoui2011} and 
those previously obtained with periodic boundary conditions in \cite{Shi2012a}. 
The membrane tank-treading velocity (scaled by $\gamma R_0/2$)
is also in good agreement with the results in \cite{Kaoui2011,Shi2012a}.
The results show that there
is no significant difference when having the Dirichlet boundary conditions
on $\Gamma$ with the length $L=112$ and $80 \ \rm{\mu m}$ or the conditions 
(\ref{eqn:2.3a}) and (\ref{eqn:2.3b}) on the boundary of the shorter domain.

\subsection{Multi-cell aggregation in a capillary behind a moving interface}\label{sec.3.2}

For the cases involving a moving interface in a capillary, we have considered the one moving to the right 
with constant speed $U$ to mimic the motion of the RBCs behind a meniscus in a capillary.
Then the associated boundary condition in (\ref{eqn:2.3a}) on $\Gamma_d$ is ${\bf g}_0={\bf 0}$ on
$\Gamma_1\cup\Gamma_3$ and ${\bf g}_0=(U,0)^t$ on $\Gamma_2$   and the boundary condition (\ref{eqn:2.3b}) 
is satisfied on $\Gamma_4$. We have kept all the related parameters the same except the following. We have 
first considered the case of 12 cells  of swelling ratio $s^*$=0.481 in a capillary of the height 
$10 \rm{\mu m}$. The computational domain $\Omega$ is $80 \rm{\mu m }\times 10 \rm{\mu m}$. 
The interface speed is $U=8/3$ cm/s. 
The cells in the center of the channel move faster than those next to the top and bottom walls do due to fact that the 
velocity field behaves like Poiseuille flow as the fluid flow is away from the interface and the speed of the 
interface is slower than the velocity of the fluid flow in the channel central region away from the interface 
(see the velocity field in Figures \ref{fig.5} and \ref{fig.6}). For the cells moving away from the interface, 
they move back to the central region of the channel due to the lateral migration of the cells in a flow field 
like the Poiseuille flow and then move toward the interface. Thus the cells form a slug behind the moving 
interface and move with the interface as in Figure \ref{fig.5}.  
For the case of 68 cells  of swelling ratio $s^*$=0.481 in a capillary 
of the height $20 \rm{\mu m}$, we have considered the computational domain 
$\Omega=160 \rm{\mu m }\times 20 \rm{\mu m}$. 
The interface speed is $U=8/3$ cm/s. These 68 cells behave similarly behind the 
moving interface like the motion of the 12 cells considered in the previous case. But it is much clearly for 
us to see that the cells in the channel central region move faster to the right due to the relatively faster 
flow field. Then the cells are piled up behind the interface and move with the interface in Figure \ref{fig.6}.

\section{Conclusions}

In summary, we have developed computational modeling and methodologies for simulating the motion 
of many RBCs in a capillary behind a moving interface in this paper.  
The methodology is based on an immersed boundary method and the skeleton structure of the red blood 
cell (RBC) membrane is modeled as a spring network.  The computational domain is moving with either a designated
RBC or an interface in an infinitely long two-dimensional channel with an undisturbed flow field in
front of the domain. The tanking-treading and the inclination angle of a cell in a simple
shear flow are briefly discussed for the validation purpose. The results of the
motion of red blood cells behind a moving interface in a capillary show that
the RBCs with higher velocity than the interface speed form a concentrated slug behind
the interface, which is consistent with the results in \cite{Chang2005,Chang2006}. The lateral migration
is also a key factor for the formation of a slug behind
the moving interface.

\section*{Acknowledgments}
The authors acknowledge the support of NSF (grant DMS-0914788). We acknowledge the helpful comments 
of James Feng, Ming-Chih Lai and  Sheldon X. Wang.


\begin{thebibliography}{9999}


\bibitem{Chang2005}
{\sc H.-C. Chang, R. Zhou},
 {\em Capillary penetration failure of blood suspensions},
 J. Colloid Interface Sci. 287 (2005), pp. 647--656.

\bibitem{Chang2006}
{\sc R. Zhou, J. Gordon, A.F. Palmer, H.-C. Chang},
 {\em Role of erythrocyte deformability during capillary wetting},
 Biotechnology and Bioengineering 93 (2006), pp. 201-–211.


\bibitem{Cristini}
{\sc V. Cristini, G.S. Kassab},
 {\em Computer odeling of red blood cell rheology in the microcirculation: a brief overview.},
 Ann. Biomed. Eng. 33 (2005), pp. 1724--1727.




\bibitem{Pozrikidis2003}
{\sc C. Pozrikidis},
 {\em Modeling and simulation of capsules and biological cells}. 
 Chapman \& Hall/CRC: Boca Raton, 2003.


\bibitem{Peskin1977}
{\sc C.S. Peskin},
 {\em Numerical analysis of blood flow in the heart},
 J.  Comput. Phys., 25 (1977), pp. 220--252.

\bibitem{Peskin2002}
{\sc C.S. Peskin},
 {\em The immersed boundary method},
 Acta Numer., 11 (2002), pp. 479--517.

\bibitem{Peskin1980}
{\sc C.S. Peskin, D.M. McQueen},
 {\em Modeling prosthetic heart valves for numerical analysis of blood flow in the heart},
 J. Comput. Phys., 37 (1980), pp. 113-–32.



\bibitem{Eggleton1998}
{\sc C. Eggleton, A. Popel},
 {\em Large deformation of red blood cell ghosts in a simple shear flow},
 Phys.  Fluids, 10 (1998), pp. 1834--1845.

\bibitem{Bagchi2005}
{\sc P. Bagchi, P. Johnson, A. Popel},
 {\em Computational Fluid Dynamic Simulation of Aggregation of Deformable Cells  in a Shear Flow},
 J. Biomech. Eng., 127 (2005), pp. 1070--1080.

\bibitem{Bagchi2007}
{\sc P. Bagchi},
 {\em Mesoscale simulation of blood flow in small vessels},
 Biophys. J., 92 (2007), pp. 1858--1877.

\bibitem{Zhang2009}
{\sc J. Zhang, J. Johnson, A.S. Popel},
 {\em Effects of erythrocyte deformability and aggregation on the cell free layer and apparent viscosity of microscopic blood flows},
 Microvasc. Res., 77 (2009), pp. 265--272.

\bibitem{Fogelson2010}
{\sc L.M. Crowl, A.L. Fogelson},
 {\em Computational model of whole blood exhibiting lateral platelet motion induced by red blood cells},
 Int. J. Numer. Meth. Biomed. Engng., 26 (2010), pp. 471--487.

\bibitem{Kaoui2011}
{\sc B. Kaoui, J. Harting, C. Misbah},
 {\em Two-dimensional vesicle dynamics under shear flow: Effect of confinement},
 Phys. Rev. E, 83 (2011), 066319.

\bibitem{Kim2012}
{\sc Y. Kim, M.-C. Lai},
 {\em Numerical study of viscosity and inertial effects on tank-treading and tumbling motions of vesicles under shear flow},
 Phys. Rev. E, 86 (2012), 066321.



\bibitem{Shi2012a}
{\sc L. Shi, T.-W. Pan, R. Glowinski},
 {\em Deformation of a single blood cell in bounded Poiseuille flows},
 Phys. Rev. E, 85 (2012), 016307. 

\bibitem{Shi2012b}
{\sc L. Shi, T.-W. Pan, R. Glowinski},
 {\em Lateral migration and equilibrium shape and position of a single red blood cell in bounded Poiseuille flows}, Phys. Rev. E, 86 (2012), 056308.  

\bibitem{Shi2012c}
{\sc L. Shi, T.-W. Pan, R. Glowinski},
 {\em Numerical simulation of lateral migration of red blood cells in Poiseuille flows},
 Int. J. Numer. Methods Fluids, 68 (2012), pp. 1393--1408.

\bibitem{Tsubota2006}
{\sc K. Tsubota, S. Wada, T. Yamaguchi},
 {\em Simulation study on effects of hematocrit on blood flow properties using particle method},
 J. Biomech. Sci. Eng., 1 (2006), pp. 159--170.


\bibitem{Pan2009b}
{\sc T. Wang, T.-W. Pan, Z. Xing, R. Glowinski},
 {\em Numerical simulation of rheology of red blood cell rouleaux in microchannels},
 Phys. Rev. E, 79 (2009), 041916.


\bibitem{Alexeev2006}
{\sc A. Alexeev, R. Verberg, A.C. Balazs},
{\it Modeling the interactions between deformable capsules rolling on a compliant surface},
{Soft Matter}  2 (2006), pp. 499--509.

\bibitem{Chorin}
{\sc A.J. Chorin, T.J.R. Hughes, M.F. McCracken, J.E. Marsden},
 {\em Product formulas and numerical algorithms},
 Comm. Pure Appl. Math., 31 (1978), pp. 205--256.


\bibitem{Glowinskibook}
{\sc R. Glowinski},
 {\em Finite element methods for incompressible viscous flow},
 in Handbook of Numerical  Analysis, Vol. IX, Ciarlet PG and  Lions JL (Eds.).
North-Holland, Amsterdam (2003), pp. 7-1176.

\bibitem{Glowinskiwave}
{\sc E.J. Dean, R. Glowinski},
 {\em A wave equation approach to the numerical solution of the Navier-Stokes
equations for incompressible viscous flow},
 C.R. Acad. Sc. Paris, S\'{e}rie 1, 325 (1997), pp. 783--791.


\bibitem{Dean1998}
{\sc E.J. Dean, R. Glowinski, T.-W. Pan},
 {\em A wave equation approach to the numerical simulation of incompressible
viscous fluid flow modeled by the Navier–Stokes equations},
 in Mathematical and Numerical Aspects of Wave Propagation, De Santo JA (Ed.).
SIAM: Philadelphia(1998), pp. 65--74.


\bibitem{Hu1992}
{\sc H.H. Hu, D.D. Joseph, M.J. Crochet},
 {\em Direct simulation of fluid particle motions},
 Theoret. Comput. Fluid Dynamics 3 (1992), pp. 285--306.

\bibitem{Pan2002c}
{\sc T.-W. Pan, R. Glowinski, G.P. Galdi},
 {\em Direct simulation of the motion of a settling ellipsoid in Newtonian fluid},
 J. Comput. Applied Math., 149 (2002), pp. 71--82. 





\end{thebibliography}
\end{document}